\documentclass[
prd,
reprint,
superscriptaddress,
longbibliography
]{revtex4-1}

\usepackage{amsmath,amssymb}
\usepackage{verbatim}
\usepackage{graphicx}
\usepackage{hyperref}
\usepackage{color}
\usepackage{mathrsfs}
\usepackage{slashed}

\usepackage{graphicx}% Include figure files
\usepackage{dcolumn}% Align table columns on decimal point
\usepackage{bm}% bold math

\usepackage{tikz}
\usetikzlibrary{arrows.meta,positioning}
\usepackage{float}

\usepackage[utf8]{inputenc}

\newcommand{\be}{\begin{equation}}
	\newcommand{\ee}{\end{equation}}
\newcommand{\bea}{\begin{eqnarray}}
	\newcommand{\eea}{\end{eqnarray}}

\newcommand{\ba}{\begin{array}}
	\newcommand{\ea}{\end{array}}

\def\double #1{#1{\hbox{\kern-2pt $#1$}}}

\newcommand{\bsubeq}{\begin{subequations}}
	\newcommand{\esubeq}{\end{subequations}}

\begin{document}
	
	\title{Supersymmetric Twisted Carroll Theories }

		\author{Osman Ergec}
	\email{ergec24@itu.edu.tr}
	\affiliation{Department of Physics,
		Istanbul Technical University,
		Maslak 34469 Istanbul,
		T\"urkiye}

	\date{\today}
	
	%	\preprint{}

\begin{abstract}
We study the recently discussed \cite{Zorba_2025,bulunur2026twistedoriginmagneticcarroll} $2+1$-dimensional Hull-type twisted $\mathcal{N}=2$ Carroll superalgebra. This structure admits both electric- and magnetic-type realizations at the level of the supersymmetry transformations. At the Lagrangian level, magnetic Carroll theories describe fields that propagate in space while remaining constrained in time, and vice versa for electric theories. We first realize this symmetry algebra through the dimensional reduction of a $2+2$-dimensional $\mathcal{N}=(1,1)$ parent algebra \cite{Bergshoeff_1992}, originally discussed in the context of self-dual supergravity. We further find that consistent reduction of this higher-dimensional algebra admit infinite dimensional lifts to left- and right-chiral super $\mathrm{BMS}_4$-type algebras, which can be combined by restoring the $R$-symmetry leading to the type I-I magnetic super- $\mathrm{BMS}_4$ \cite{Zheng:2025rfe, bulunur2026twistedoriginmagneticcarroll}. We then consistently construct magnetic off-shell scalar and Abelian Yang–Mills theories, and comment on on-shell theories. Finally, in the appendix, we explain how an ordinary supersymmetric theory can be analytically continued to its twisted counterpart.
\end{abstract}

	%	\pacs{??? ... ???}
	
	\maketitle
	\allowdisplaybreaks

%%%%%%%%%%%%%%%%%%%%%%%%%%%%%%%%%%%%%%%%%%%%%%%%%%%%%%%%%%%%%%%%%%%%%%%%%%%%%%%%
	 \textit{Introduction}--  The term ultra-relativistic limit for $c \rightarrow 0$ can be misleading, since it is often understood as the usual ultra-relativistic regime $v \rightarrow c$ at fixed $c$ \cite{deboer2021carrollsymmetrydarkenergy}. Although both limits yield the same leading contribution to the energy, $E \simeq pc$, they represent distinct limiting procedures. By contrast, the Carroll limit is a group contraction---an irreversible operation on the theory---in which $c$ itself is taken to a singular limit. Despite their different geometric interpretations, from an algebraic point of view, the two singular contractions, $c\rightarrow0$ and $c\rightarrow\infty$, of the $U(1)$-extended Poincaré algebra lead to the same algebraic structure, namely the Bargmann algebra, which contains the Carroll algebra as a subalgebra. This procedure consistently suppresses certain aspects of the parent theory and isolates a distinct sector with its own dynamics. Although field theories cannot directly see this subalgebra relation, the contraction is realized directly at the level of their dynamics and field-theoretic description, where additional symmetry enhancement may also arise. Therefore, this symmetry class may have a wide range of applications, ranging from open/closed-string duality \cite{Blair:2023noj,Blair:2024aqz,Blair:2025nno} to generalized symmetries \cite{Bidussi:2021nmp,Perez:2023uwt,Baig:2023yaz,Kasikci:2023tvs,Huang:2023zhp,Figueroa-OFarrill:2023vbj,Kasikci:2023zdn,Ecker:2024czx,Hartong:2024hvs}, and may also provide a route toward flat-space holography \cite{Bagchi:2016bcd,Donnay:2022aba,Donnay:2022wvx,Saha:2023hsl,Nguyen:2023vfz,Ruzziconi:2026bix}.  
    
    In particular, we regard flat space as arising from an appropriate limit
of an AdS geometry, which motivates the flat-space holographic viewpoint
and aligns naturally with the generalized-symmetry perspective. In
addition, motivated in part by the discovery of gravitational waves,
Carrollian symmetry \cite{AIHPA_1965__3_1_1_0} is important because it
is nontrivially related to the asymptotic symmetry structures expected
to arise in quantum gravity \cite{Strominger:2017zoo}. This connection
has been clarified primarily over the last decade
\cite{Duval_2014a,Duval_2014b}, notably through its lift to the global
$\mathrm{BMS}$ algebra \cite{Bondi1962GravitationalWI}, which is
expected to encode quantum-gravitational symmetries. The global
symmetries of the bulk are realized in the dual gauge theory as
spacetime and internal symmetries.

Here, we study a class of Carrollian symmetry first proposed in
\cite{Zorba_2025} and further developed in
\cite{bulunur2026twistedoriginmagneticcarroll}, whose supersymmetric extension accommodates
both spatial and temporal diffeomorphisms, in contrast to the
exclusively temporal diffeomorphisms discussed in
\cite{Bergshoeff_2016p}. In the present work, we derive this symmetry
from a $2+2$ dimensional $\mathcal{N}=(1,1)$ parent superconformal
algebra originally discussed in the context of self-dual supergravity
\cite{Bergshoeff_1992}. This reduction leads to the Type A and Type B
Carrollian superalgebras and clarifies their relation to the left- and
right-chiral super-$\mathrm{BMS}_4$ structures. We also construct scalar
and vector multiplets realizing this symmetry, including an off-shell magnetic scalar model with a supersymmetric multiplier sector and a consistent magnetic
truncation of the hybrid Yang--Mills multiplet. For off-shell theories,
the Carroll contraction can be studied directly through the
transformation rules and Lagrangian. For on-shell theories, however,
there are additional constraints: the singular limit must also be
consistent with the equations of motion. We begin by introducing the
Hull-type twisted $\mathcal{N}=2$ Carroll algebra in $2+1$ dimensions.
\begin{equation} \label{SuperCarrollAlgebra}
\begin{aligned}
&[J, P_a] = \varepsilon_{ab} P^b, \ \ \
 [J, C_a] = \varepsilon_{ab} C^b, \ \ \ [C_a, P_b] = -\,\varepsilon_{ab} H,  \\[4pt]
& [C_a, \mathcal{Q}_\alpha^{+}] = \tfrac12 (\gamma_a \mathcal{Q}^{-})_\alpha,  \ \ \ [J, \mathcal{Q}_\alpha^{\pm}] 
  = \tfrac12 (\gamma_0 \mathcal{Q}^{\pm})_\alpha, \\[4pt]
&\{ \mathcal{Q}_\alpha^{+}, \mathcal{Q}_\beta^{+} \} 
  = (\gamma^a C^{-1})_{\alpha\beta} P_a, \ \  \
\{ \mathcal{Q}_\alpha^{+}, \mathcal{Q}_\beta^{-} \} 
  = (\gamma^0 C^{-1})_{\alpha\beta} H, \\[4pt]
\end{aligned}
\end{equation}
where $J$ is the spatial rotation generator, $P_a$ are the generators of spatial translations, $C_a$ are Carroll boosts, $H$ is the time-translation generator, and $\mathcal{Q}_\alpha^{ \pm}$ are Majorana supercharges. 

%%%%%%%%%%%%%%%%%%%%%%%%%%%%%%%%%%%%%%%%%%%%%%%%%%%%%%%%%%%%%%%%%%%%%%%%%%%%%%%%%%%%%%%%%%%%%%%%%
\textit{Reduction of the $\mathcal{N}=(1,1)$ Algebra in $2+2$ Dimensions}--
%%%%%%%%%%%%%%%%%%%%%%%%%%%%%%%%%%%%%%%%%%%%%%%%%%%%%%%%%%%%%%%%%%%%%%%%%%%%%%%%%%%%%%%%%%%%%%%%%
The structure underlying magnetic- and electric-type Carroll symmetries can
be derived from a higher dimensional parent algebra arising in self-dual
supergravity in $2+2$ dimensions \cite{Bergshoeff_1992}. We begin with the
higher dimensional superconformal algebra $\mathrm{SL}(4\mid 1)$, whose
bosonic subalgebra is
$\mathrm{SL}(4)\times\mathrm{SO}(1,1)$. Its bosonic generators consist of
the $\mathrm{SO}(2,2)$ generators $M_{AB}$, together with the translations $P_A$, conformal boosts $K_A$, the dilatation $D$, and the
$\mathrm{SO}(1,1)$ generator $A$. Using the local isomorphism
\begin{equation}
\mathrm{SO}(2,2)
\cong
\mathrm{SL}(2,\mathbb{R})\times\mathrm{SL}(2,\mathbb{R}),
\end{equation}
one can make the left- and right-chiral structure of the algebra explicit.
The bosonic part of the algebra reads
\begin{equation}
\label{2+2 Algebra1}
\begin{aligned}
\bigl[M_{AB},M_{CD}\bigr]
&=2\eta_{B[C}M_{A D]}-2\eta_{A[C}M_{B D]},
\\
\bigl[M_{AB},P_C\bigr]
&=2\eta_{[BC}P_{A]},
\qquad
\bigl[M_{AB},K_C\bigr]
=2\eta_{[BC}K_{A]},
\\
\bigl[P_A,K_B\bigr]
&=2\bigl(\eta_{AB}D-M_{AB}\bigr),
\\
\bigl[P_A,D\bigr]
&=P_A,
\qquad
\bigl[K_A,D\bigr]
=-K_A.
\end{aligned}
\end{equation}
The fermionic generators $Q_I$ and $S_I$, with
$I=\mathrm L,\mathrm R$, are pseudo-Majorana--Weyl spinors, where
$\mathrm L$ and $\mathrm R$ denote their chirality. Their mixed
commutators and fermionic anticommutators are
\begin{equation}
\label{2+2 Algebra2}
\begin{gathered}
\begin{aligned}
\{Q_{\mathrm L},Q_{\mathrm R}\}
&=-2\bigl(P_{+}\Gamma^{A}C\bigr)P_{A},
\\
\{S_{\mathrm L},S_{\mathrm R}\}
&=2\bigl(P_{+}\Gamma^{A}C\bigr)K_{A},
\\
\{S_{\mathrm L},Q_{\mathrm L}\}
&=P_{+}\bigl(2D+\Gamma^{AB}M_{AB}+A\bigr)C,
\\
\{S_{\mathrm R},Q_{\mathrm R}\}
&=P_{-}\bigl(2D+\Gamma^{AB}M_{AB}-A\bigr)C,
\end{aligned}
\\[2pt]
\begin{alignedat}{2}
[P_A,S_{\mathrm L}]
&=-\Gamma_AQ_{\mathrm R},
\qquad&
[P_A,S_{\mathrm R}]
&=-\Gamma_AQ_{\mathrm L},
\\
[K_A,Q_{\mathrm L}]
&=\Gamma_AS_{\mathrm R},
&
[K_A,Q_{\mathrm R}]
&=\Gamma_AS_{\mathrm L},
\\
[M_{AB},Q_I]
&=-\tfrac12\Gamma_{AB}Q_I,
&
[M_{AB},S_I]
&=-\tfrac12\Gamma_{AB}S_I,
\\
[D,Q_I]
&=-\tfrac12Q_I,
&
[D,S_I]
&=\tfrac12S_I,
\\
[A,Q_{\mathrm L}]
&=3Q_{\mathrm L},
&
[A,Q_{\mathrm R}]
&=-3Q_{\mathrm R},
\\
[A,S_{\mathrm L}]
&=-3S_{\mathrm L},
&
[A,S_{\mathrm R}]
&=3S_{\mathrm R}.
\end{alignedat}
\end{gathered}
\end{equation}
Here \(I=\mathrm L,\mathrm R\), and
\begin{equation}
P_{\pm}=\frac12\left(1\pm\Gamma_5\right),
\qquad
\eta_{AB}=\operatorname{diag}(-,-,+,+).
\end{equation}
with $A,B=0,1,2,3$. We denote the second timelike direction by $y$ and
decompose the parent generators according to
\begin{equation}
\begin{aligned}
&H=P_0,
\qquad
C_a=M_{0a},
\qquad
K=K_0,
\\
&Z=P_y,
\qquad
\widetilde Z=K_y,
\qquad
B_\mu=M_{\mu y},
\\
&B=B_0,
\qquad
\Xi_Q=P_+\Gamma_yQ_{\mathrm R},
\qquad
\Xi_S=P_+\Gamma_yS_{\mathrm R}.
\end{aligned}
\end{equation}
Here $\mu=0,1,2$ and $a=1,2$. The reduced gamma matrices and
charge-conjugation matrix are defined by
\begin{equation}
\gamma_\mu=P_+\Gamma_\mu\Gamma_y,
\qquad
C_3=P_+C.
\end{equation}
All projected spinors and matrices act on the reduced two-component
spinor space. We then introduce the three dimensional supercharge basis
\begin{equation}
\begin{gathered}
\begin{aligned}
\widehat Q^\pm_\alpha
&=
\left(\mathcal{A}_\pm Q_{\mathrm L}\right)_\alpha
+\left(\mathcal{A}_\mp\Xi_Q\right)_\alpha,
\\
\widehat S^\pm_\alpha
&=
\left(\mathcal{A}_\pm S_{\mathrm L}\right)_\alpha
+\left(\mathcal{A}_\mp\Xi_S\right)_\alpha,
\end{aligned}
\\[3pt]
\mathcal{A}_\pm=\frac12\left(1\pm\gamma_0\right),
\qquad
\gamma_0^2=-1.
\end{gathered}
\end{equation}
$\mathcal A_\pm$ define an invertible real change of basis. For notational simplicity, we henceforth suppress the hats and denote the redefined three dimensional charges by $Q^\pm$ and $S^\pm$. The
superscripts $\pm$ therefore label the reduced charge basis rather than
the chirality of the parent charges.

In this basis, the Carroll contraction is defined by the rescalings
\begin{equation}
\label{Carrollian scaling of 2+2 d superconformalalgebra}
\begin{aligned}
&H\rightarrow c^{-1}H,
\qquad
C_a\rightarrow c^{-1}C_a,
\qquad
K\rightarrow c^{-1}K,
\\
&B\rightarrow c^{-1}B,
\qquad
A\rightarrow c^{-1}A,
\qquad
Q^-\rightarrow c^{-1}Q^-,
\\
&S^-\rightarrow c^{-1}S^-.
\end{aligned}
\end{equation}
All generators not displayed above are held fixed. Taking $c\rightarrow0$
then yields the $\mathcal{N}=(1,1)$ superconformal Carroll algebra in
$2+2$ dimensions.

This algebra admits
two consistent truncations along the unphysical timelike direction,
leading to two classes of $2+1$ dimensional algebras, which we refer to
as Type A and Type B, as shown in Fig. \ref{fig:tsca-diagram2}.\begin{figure}[H]
\centering
\begin{tikzpicture}[
  scale=0.95,
  transform shape,
  box/.style={
    draw,
    rounded corners,
    inner sep=3pt,
    align=center,
    font=\scriptsize\bfseries
  },
  bmsbox/.style={
    draw,
    rounded corners,
    inner sep=3pt,
    align=center,
    font=\scriptsize,
    text width=4.2cm
  },
  arrow/.style={
    -{Latex[length=1.4mm,width=1mm]},
    line width=0.55pt,
    shorten >=4pt,
    shorten <=4pt
  },
  lab/.style={
    font=\tiny\bfseries,
    inner sep=1pt,
    fill=white
  }
]

\node[box] (scca22) {$2+2$\\$\mathcal{N}=(1,1)$\\SCCA};

\node[box, below=15mm of scca22, xshift=-23mm] (typeA)
{$2+1$\\Type A\\$(M_{ab},C_a,P_a,H,Z,D,A,Q^\pm)$};

\node[box, below=15mm of scca22, xshift=23mm] (typeB)
{$2+1$\\Type B\\$(M_{ab},C_a,K_a,K,\tilde Z,D,A,S^\pm)$};

% shorter downward arrows by reducing vertical distance
\node[bmsbox, below=8mm of typeA] (leftbms)
{Left chiral $BMS_4$\\
$\left\{L_n,\bar{L}_n,M_{r,s},G_\rho,N_{k,\rho}\right\}$};

\node[bmsbox, below=8mm of typeB] (rightbms)
{Right chiral $BMS_4$\\
$\left\{L_n,\bar{L}_n,M_{r,s},\bar{G}_\rho,\bar{N}_{k,\rho}\right\}$};

\draw[arrow] (scca22) -- (typeA)
  node[midway, lab, above, sloped] {Trunc.};

\draw[arrow] (scca22) -- (typeB)
  node[midway, lab, above, sloped] {Trunc.};

\draw[arrow] (typeA) -- (leftbms);
\draw[arrow] (typeB) -- (rightbms);

\end{tikzpicture}
\caption{The $2+2$ (super)conformal Carroll algebra gives rise to the Type A and Type B Carrollian algebras in $2+1$ dimensions. They admit infinite dimensional lift to the left- and right-chiral super-$BMS_4$ structures.}
\label{fig:tsca-diagram2}
\end{figure}
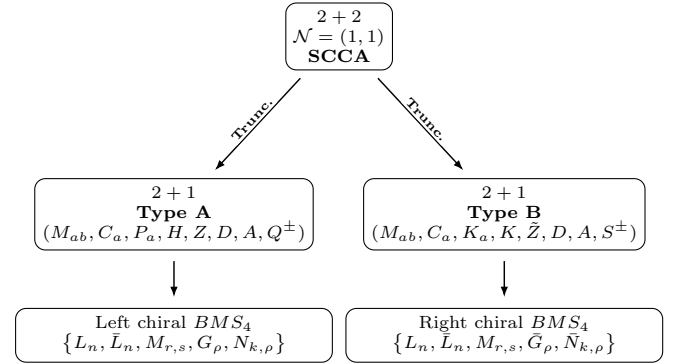
\begin{equation}  \label{Type A}
  \mathrm{Type~A}:
  \left(M_{ab}, C_a, P_a, H, Z, D, A, {Q}^{\pm}\right),
\end{equation}
with spatial indices $a,b,c,d=1,2$, realizing twisted $\mathcal{N}=2$ supersymmetry in $2+1$ dimensions. Upon turning off the charges $A$, $Z$, and $D$, and writing the generators in the dual Lorentz basis, one recovers the algebra in Eq.~\eqref{SuperCarrollAlgebra}. The Type A algebra can equivalently be obtained by first reducing the parent theory along the timelike direction to $2+1$ dimensions and then performing the Carroll contraction. Its nonvanishing (anti)commutators are given in the bosonic sector by
\begin{equation}
\label{eq:truncated-lorentz}
\begin{aligned}
\left[M_{ab},M_{cd}\right]
&=4\,\delta_{[a[c}M_{d]b]},\\
\left[M_{ab},X_c\right]
&=2\,\delta_{c[b}X_{a]},
\ \ X_c\in\{C_c,P_c\},\ \
\left[C_a,P_b\right]
=\delta_{ab}H,\\
\left[P_a,D\right]
&=P_a,
\qquad
\left[H,D\right]
=H,
\qquad
\left[Z,D\right]
=Z,
\end{aligned}
\end{equation}
mixed sector
\begin{equation}
\label{eq:truncated-mixed}
\begin{aligned}
[M_{ab},{Q^{\pm}}]
&=-\tfrac12\gamma_{ab} {Q^{\pm}},
\quad
[C_a,{Q^{+}}]
=-\tfrac12\gamma_{0a} {Q^{-}},
\\
[D,{Q^{\pm}}]
&=-\tfrac12 {Q^{\pm}},
\quad
[A,{Q^{+}}]
=3\gamma_0{Q^{-}},
\end{aligned}
\end{equation}
and fermionic sector
\begin{equation}
\label{eq:truncated-fermionic}
\begin{aligned}
\{{Q}^{+},{Q}^{+}\}
=2\gamma^{a}C_{3}P_a-2\gamma^{0}C_{3}Z,
\ \ 
\{{Q}^{+},{Q}^{-}\}
=2\gamma^{0}C_{3}H.
\end{aligned}
\end{equation}

In addition, there is a second, essentially similar truncation leading
to the Carroll-type algebra
\begin{equation}
  \mathrm{Type~B}:
  \left(M_{ab},C_a,K_a,K,\widetilde{Z},D,A,S^\pm\right).
\end{equation}
The Type B algebra is related to Type A by the conformal exchange
\begin{equation}
\begin{aligned}
P_a&\leftrightarrow-K_a, & H&\leftrightarrow-K,\\
Z&\leftrightarrow-\widetilde{Z}, &
Q^\pm&\leftrightarrow S^\pm,\\
D&\leftrightarrow-D, & A&\leftrightarrow-A,
\end{aligned}
\end{equation}
with $M_{ab}$ and $C_a$ left unchanged. However, the full twisted
(super)conformal $\mathcal{N}=2$ algebra in $2+1$ dimensions cannot be
obtained by timelike reduction of the undeformed $2+2$ dimensional
algebra alone, even though the Type A and Type B Carrollian
superalgebras exist. Instead, one must introduce a deformation in the
sector coupling the two algebras, namely in the mixed anticommutator
$\{Q^\pm,S^\mp\}$, and fix it by imposing the Jacobi identities.
Alternatively, one may begin with either the Type A or Type B
Carrollian superalgebra and construct the corresponding twisted
superconformal Carroll extension. In addition, once the $R$-symmetry
and central charges are turned off, the Type A and Type B algebras
admit lifts to the left- and right-chiral
super-$\mathrm{BMS}_4$-type structures, respectively, as recently
discussed in \cite{Zheng:2025cuw}.

\newcommand{\chiralsetL}{\bigl\{\,L_n,\bar L_n,M_{r,s},G_\rho,N_{k,\rho}\,\bigr\}}
\newcommand{\chiralsetR}{\bigl\{\,L_n,\bar L_n,M_{r,s},\bar G_\rho,\bar N_{k,\rho}\,\bigr\}}
\paragraph{\textbf{Left chiral.}}
\begin{equation}
\chiralsetL,
\quad n,k\in\mathbb Z,\quad \rho,r,s\in\mathbb Z+\tfrac12.
\end{equation}

\noindent\textit{Bosonic sector.}
\begin{equation}
\small
\begin{aligned}
&[L_n,L_m]=(n-m)L_{n+m}, \quad [\bar L_n,\bar L_m]=(n-m)\bar L_{n+m},\\
&[L_m,M_{r,s}]=\Big(\tfrac{m}{2}-r\Big)M_{m+r,s}, \quad
[\bar L_m,M_{r,s}]=\Big(\tfrac{m}{2}-s\Big)M_{r,m+s},\\
&[M_{r,s},M_{t,u}]=0, \quad [L_n,\bar L_m]=0.
\end{aligned}
\end{equation}

\noindent\textit{Mixed sector.}
\begin{equation}
\begin{aligned}
&[L_n,G_\rho]=\Big(\tfrac{n}{2}-\rho\Big)G_{n+\rho}, \quad [\bar L_n,G_\rho]=0,\\
&[L_n,N_{k,\rho}]=-k\,N_{n+k,\rho}, \quad
[\bar L_n,N_{k,\rho}]=\Big(\tfrac{n}{2}-\rho\Big)N_{k,n+\rho},\\
&[M_{r,s},G_\rho]=\tfrac12(\rho-r)\,N_{r+\rho,s}, \quad [M_{r,s},N_{k,\rho}]=0.
\end{aligned}
\end{equation}

\noindent\textit{Fermionic sector.}
\begin{equation}
\begin{aligned}
&\{G_\rho,G_\sigma\}=L_{\rho+\sigma}, \quad
\{G_\rho,N_{k,\sigma}\}=-\,M_{\rho+k,\sigma},\\
&\{N_{k,\rho},N_{p,\sigma}\}=0.
\end{aligned}
\end{equation} \\

\paragraph{\textbf{Right chiral.}}
\begin{equation}
\chiralsetR,
\quad n,k\in\mathbb Z,\quad \rho,r,s\in\mathbb Z+\tfrac12.
\end{equation}

\noindent\textit{Bosonic sector.}
\begin{equation}
\small
\begin{aligned}
&[L_n,L_m]=(n-m)L_{n+m}, \quad [\bar L_n,\bar L_m]=(n-m)\bar L_{n+m},\\
&[L_m,M_{r,s}]=\Big(\tfrac{m}{2}-r\Big)M_{m+r,s}, \quad
[\bar L_m,M_{r,s}]=\Big(\tfrac{m}{2}-s\Big)M_{r,m+s},\\
&[M_{r,s},M_{t,u}]=0, \quad [L_n,\bar L_m]=0.
\end{aligned}
\end{equation}

\noindent\textit{Mixed sector.}
\begin{equation}
\begin{aligned}
&[\bar L_n,\bar G_\rho]=\Big(\tfrac{n}{2}-\rho\Big)\bar G_{n+\rho}, \quad [L_n,\bar G_\rho]=0,\\
&[\bar L_n,\bar N_{k,\rho}]=-k\,\bar N_{n+k,\rho}, \quad
[L_n,\bar N_{k,\rho}]=\Big(\tfrac{n}{2}-\rho\Big)\bar N_{k,n+\rho},\\
&[M_{r,s},\bar G_\rho]=\tfrac12(\rho-s)\,\bar N_{s+\rho,r}, \quad [M_{r,s},\bar N_{k,\rho}]=0.
\end{aligned}
\end{equation}

\noindent\textit{Fermionic sector.}
\begin{equation}
\begin{aligned}
&\{\bar G_\rho,\bar G_\sigma\}=-\,\bar L_{\rho+\sigma}, \quad
\{\bar G_\rho,\bar N_{k,\sigma}\}= M_{\sigma,\rho+k},\\
&\{\bar N_{k,\rho},\bar N_{p,\sigma}\}=0.
\end{aligned}
\end{equation}
These chiral Type $A/B$ structures can then be glued together via the \(R\)-symmetry, leading to the magnetic super-\(BMS_4\) algebra recently discussed in \cite{Zheng:2025rfe, bulunur2026twistedoriginmagneticcarroll}.

%%%%%%%%%%%%%%%%%%%%%%%%%%%%%%%%%%%%%%%%%%%%%%%%%%%%%%%%%%%%%%%%%%%%%%%%%%%%%%%%
\textit{Off-Shell Magnetic Carroll Scalar Multiplet}--
%%%%%%%%%%%%%%%%%%%%%%%%%%%%%%%%%%%%%%%%%%%%%%%%%%%%%%%%%%%%%%%%%%%%%%%%%%%%%%%%
To our knowledge, there are two distinct Hull-type twisted Carrollian
scalar multiplets in $2+1$ dimensions. One of them was first obtained
using the superspace method, and details can be found in
\cite{Zorba_2025}. The same multiplet can also be derived by taking the
Carrollian limit after an appropriate field redefinition. This method
splits the multiplet into two identical pieces, up to a field
redefinition,
\begin{equation} \label{Splited Structure}
    \left(A_1,\chi_{-},F_1\right)
    \oplus
    \left(A_2,\chi_{+},F_2\right).
\end{equation}
Interestingly, this decomposed structure is preserved under the
off-shell mass deformation detailed in Appendix B,
Eq.~\eqref{app:massive-scalar}. The other Carrollian multiplet leads to
a magnetic-type Carroll scalar model,
\begin{equation} \label{configuration-space multiplet}
\begin{gathered}
\delta\phi_1
 =\bar{\epsilon}_{+}\chi_{-}
 +\bar{\epsilon}_{-}\chi_{+},
\qquad
\delta\phi_2
 =\bar{\epsilon}_{+}\gamma_0\chi_{+},
\\
\delta\chi_{+}
 =-\left(
 \gamma_0\gamma^a\partial_a\phi_2+\gamma_0F_2
 \right)\epsilon_{+}
 -2\left(\partial_0\phi_2\right)\epsilon_{-},
\\
\delta\chi_{-}
 =\left(\gamma^a\partial_a\phi_1-F_1\right)\epsilon_{+}
 +\left(
 2\gamma^0\partial_0\phi_1
 +\gamma_0\gamma^a\partial_a\phi_2+\gamma_0F_2
 \right)\epsilon_{-},
\\
\delta F_1
 =-\bar{\epsilon}_{+}\gamma^a\partial_a\chi_{-}
 -\bar{\epsilon}_{-}\gamma^a\partial_a\chi_{+}
 -2\bar{\epsilon}_{-}\gamma^0\partial_0\chi_{-},
\\
\delta F_2
 =-\bar{\epsilon}_{+}\gamma_0\gamma^a\partial_a\chi_{+}
 +2\bar{\epsilon}_{-}\partial_0\chi_{+}.
\end{gathered}
\end{equation}
These transformations close off shell according to
\begin{equation}
\begin{aligned}
\left[\delta^{(1)},\delta^{(2)}\right]\Phi
={}&-2\left(
\bar{\epsilon}_{+}^{(1)}\gamma^0\epsilon_{-}^{(2)}
+\bar{\epsilon}_{-}^{(1)}\gamma^0\epsilon_{+}^{(2)}
\right)\partial_0\Phi
\\
&-2\bar{\epsilon}_{+}^{(1)}
\gamma^a\epsilon_{+}^{(2)}\partial_a\Phi .
\end{aligned}
\end{equation}
The corresponding off-shell Lagrangian is
\begin{equation} \label{configuration-space Lagrangian}
\begin{aligned}
\mathcal{L}={}&
-\left(\partial_a\phi_2\right)^2
-2F_1\partial_0\phi_2
-2F_2\partial_0\phi_1
+F_2^2
+\bar{\chi}_{+}\gamma^a\partial_a\chi_{+}
\\
&+\bar{\chi}_{-}\gamma^0\partial_0\chi_{+}
+\bar{\chi}_{+}\gamma^0\partial_0\chi_{-}
-2m\phi_2F_2
+m\bar{\chi}_{+}\chi_{+}.
\end{aligned}
\end{equation}
The massive contribution is introduced through a superpotential term.
To make the magnetic structure manifest, we separate the part
containing only the fields $(\phi_2,\chi_{+},F_2)$,
\begin{equation}
\begin{aligned}
\mathcal{L}_{\mathrm{mag}}={}&
-\left(\partial_a\phi_2\right)^2+F_2^2
+\bar{\chi}_{+}\gamma^a\partial_a\chi_{+}
\\
&-2m\phi_2F_2+m\bar{\chi}_{+}\chi_{+}.
\end{aligned}
\end{equation}
Using integration by parts and the Majorana flip identities, the full
Lagrangian can be written, up to a total derivative, as
\begin{equation}
\mathcal{L}
\doteq
\mathcal{L}_{\mathrm{mag}}
-2F_1\partial_0\phi_2
+2\phi_1\partial_0F_2
+2\bar{\chi}_{-}\gamma^0\partial_0\chi_{+}.
\end{equation}
Thus, $(F_1,\phi_1,\chi_{-})$ form a multiplier sector required by the full supersymmetry.
Their equations of motion impose the magnetic Carroll conditions
\begin{equation} \label{magnetic-constraints}
\partial_0\phi_2=0,
\qquad
\partial_0\chi_{+}=0,
\qquad
\partial_0F_2=0.
\end{equation}
These conditions define a supersymmetry-invariant constraint surface.
Indeed,
\begin{equation}
\begin{aligned}
\delta\left(\partial_0\phi_2\right)
={}&\bar{\epsilon}_{+}\gamma_0\partial_0\chi_{+},
\\
\delta\left(\partial_0\chi_{+}\right)
={}&-\left(
\gamma_0\gamma^a\partial_a\partial_0\phi_2
+\gamma_0\partial_0F_2
\right)\epsilon_{+}
-2\partial_0^2\phi_2\,\epsilon_{-},
\\
\delta\left(\partial_0F_2\right)
={}&-\bar{\epsilon}_{+}\gamma_0\gamma^a
\partial_a\partial_0\chi_{+}
+2\bar{\epsilon}_{-}\partial_0^2\chi_{+},
\end{aligned}
\end{equation}
which vanish upon imposing Eq.~\eqref{magnetic-constraints}. Both
$\epsilon_{+}$ and $\epsilon_{-}$ are retained, so no supercharge is
truncated. Although the $\epsilon_{-}$ transformations act trivially on
$(\phi_2,\chi_{+},F_2)$ on the magnetic constraint surface, they remain
nontrivial on the multiplier sector. The complete action therefore
provides an off-shell $\mathcal{N}=2$ realization whose equations of
motion restrict the magnetic multiplet to be time independent.

%%%%%%%%%%%%%%%%%%%%%%%%%%%%%%%%%%%%%%%%%%%%%%%%%%%%%%%%%%%%%%%%%%%%%%%%%%%%%%%%
\textit{On-Shell Consistency of Carroll Contractions}--
%%%%%%%%%%%%%%%%%%%%%%%%%%%%%%%%%%%%%%%%%%%%%%%%%%%%%%%%%%%%%%%%%%%%%%%%%%%%%%%%
The preceding construction is off shell. For on-shell theories, however,
contracting the transformation rules alone is not sufficient: the contraction
must also be implemented consistently at the level of the equations of motion
used to establish closure. To illustrate this point, consider the
$\mathcal{N}=1$ self-dual massive vector multiplet in $2+1$ dimensions
constructed in \cite{Townsend:1983xs}. Although this example does not realize
the Hull-type twisted $\mathcal{N}=2$ algebra, it isolates the additional
consistency condition that arises when contracting an on-shell multiplet. At
first sight, the relativistic theory appears to admit two candidate
Carrollian scalings. The first is
\begin{equation}
\label{eq:self-dual-scaling-I}
\begin{aligned}
\partial_0&\rightarrow c^{-1}\partial_0,
&\qquad
m&\rightarrow c^{-1/2}m,
&\qquad
\epsilon&\rightarrow c\,\epsilon,
\\
A_0&\rightarrow cA_0,
&
A_a&\rightarrow cA_a,
&
\lambda&\rightarrow c^{1/2}\lambda,
\end{aligned}
\end{equation}
whereas the second is
\begin{equation}
\label{eq:self-dual-scaling-II}
\begin{aligned}
\partial_0&\rightarrow c^{-1}\partial_0,
&\qquad
m&\rightarrow cm,
&\qquad
\epsilon&\rightarrow c^{1/2}\epsilon,
\\
A_0&\rightarrow c^{-1}A_0,
&
\lambda&\rightarrow c^{1/2}\lambda.
\end{aligned}
\end{equation}
All quantities not displayed in these expressions are held fixed. At the
algebraic level, both candidate scalings give the Carrollian supersymmetry
algebra
\begin{equation}
\label{eq:self-dual-Carroll-algebra}
\{Q_\alpha,Q_\beta\}
=
\frac{1}{2}
(\gamma^0C)_{\alpha\beta}H.
\end{equation}
The first scaling, however, is incompatible with the equations of motion
required for the on-shell closure of the multiplet and produces divergent
terms in the Carrollian limit. By contrast, the second scaling is consistent
with the equations of motion and therefore yields a well-defined Carrollian
theory. The resulting Lagrangian is
\begin{equation}
\label{eq:self-dual-Carroll-Lagrangian}
\mathcal{L}
=
\frac{1}{2}m^2A_0^2
+m\varepsilon^{ab}A_0\partial_aA_b
-\frac{m}{2}\varepsilon^{ab}A_a\partial_0{A}_b
-\frac{1}{2}\bar{\lambda}\gamma^0\partial_0\lambda.
\end{equation}
The corresponding supersymmetry
transformations are
\begin{equation}
\label{eq:self-dual-Carroll-transformations}
\begin{aligned}
\delta A_0
&=-\frac{1}{2m}\bar{\epsilon}\,\partial_0\lambda,
\\
\delta A_a
&=-\frac{1}{2m}\bar{\epsilon}\,\partial_a\lambda,
\\
\delta\lambda
&=-\frac{1}{2}m\gamma^0A_0\epsilon.
\end{aligned}
\end{equation}
The equations of motion take the form
\begin{equation}
\label{eq:self-dual-Carroll-equations}
\begin{aligned}
mA_0+\varepsilon^{ab}\partial_aA_b&=0,
\\
\partial_0{A}_a-\partial_aA_0&=0,
\\
\partial_0{\lambda}&=0.
\end{aligned}
\end{equation}
Using these equations, the supersymmetry transformations close on shell
according to
\begin{equation}
\label{eq:self-dual-Carroll-closure}
\left[\delta^{(1)},\delta^{(2)}\right]\Phi
=
\frac{1}{2}
\left(
\bar{\epsilon}^{(2)}\gamma^0\epsilon^{(1)}
\right)
\partial_0\Phi .
\end{equation}
This example shows that on-shell closure is not automatically preserved
under contraction. While off-shell closure follows directly from the
transformation rules, on-shell closure also depends on the equations of
motion. One must therefore require that the contraction of the parent
equations of motion agrees with the equations of motion derived from the
contracted action and used to establish the on-shell closure.

%%%%%%%%%%%%%%%%%%%%%%%%%%%%%%%%%%%%%%%%%%%%%%%%%%%%%%%%%%%%%%%%%%%%%%%%%%%%%%%%
\textit{Hybrid Carrollian $\mathcal{N}=(1,1)$ Yang--Mills Theory
and Its Magnetic-Type Truncation}---
%%%%%%%%%%%%%%%%%%%%%%%%%%%%%%%%%%%%%%%%%%%%%%%%%%%%%%%%%%%%%%%%%%%%%%%%%%%%%%%%
A particularly instructive example is provided by the four-dimensional
$\mathcal{N}=(1,1)$ Abelian vector multiplet in split signature $(2,2)$.
We decompose the coordinates as $x^M=(x^0,x^a,y)$, where $a=1,2$ label
the spatial directions, $x^0$ is the physical time coordinate, and $y$
denotes the second timelike direction. We identify
\begin{equation}
A_y\equiv\sigma,
\qquad
F_{MN}=2\partial_{[M}A_{N]}.
\end{equation}
We retain the $y$ dependence below; a strict dimensional reduction to
$(2+1)$ dimensions is obtained by imposing $\partial_y=0$.

For later convenience, the relativistic Lagrangian is decomposed as
\begin{equation}
\mathcal{L}_{\text{rel}}
=
\mathcal{L}_{\text{hyb}}
+
\mathcal{L}_{\text{mag}},
\end{equation}
where
\begin{equation}
\begin{aligned}
\mathcal{L}_{\text{hyb}}
={}&
\frac{1}{2}F_{0a}^{2}
-\frac{1}{2}F_{0y}^{2}
+\bar{\lambda}_{-}\gamma^0\partial_0\lambda_{+}
+\frac{1}{2}\bar{\lambda}_{-}\gamma^a\partial_a\lambda_{-}
\\
&-\frac{1}{2}\bar{\lambda}_{-}\gamma_0\partial_y\lambda_{-}
-\frac{1}{2}D^2,
\\[2pt]
\mathcal{L}_{\text{mag}}
={}&
-\frac{1}{4}F_{ab}F_{ab}
+\frac{1}{2}F_{ay}F_{ay}
+\frac{1}{2}\bar{\lambda}_{+}
\left(\gamma^a\partial_a+\gamma_0\partial_y\right)\lambda_{+}.
\end{aligned}
\end{equation}
Here, the labels $\mathcal{L}_{\text{hyb}}$ and
$\mathcal{L}_{\text{mag}}$ refer to the hybrid and magnetic sectors
with respect to the physical Carroll time $x^0$.

The theory admits two distinct Carrollian scaling prescriptions. The
first is
\begin{equation}
\begin{aligned}
&\partial_0\rightarrow c^{-1}\partial_0,
\qquad
\sigma\rightarrow c\sigma,
\qquad
A_a\rightarrow cA_a,
\\
&\lambda_{+}\rightarrow c\lambda_{+},
\qquad
\epsilon_{-}\rightarrow c\epsilon_{-},
\end{aligned}
\end{equation}
whereas the second is
\begin{equation}
\begin{aligned}
&\partial_0\rightarrow c^{-1}\partial_0,
\qquad
D\rightarrow c^{-1}D,
\qquad
A_0\rightarrow c^{-1}A_0,
\\
&\lambda_{-}\rightarrow c^{-1}\lambda_{-},
\qquad
\epsilon_{-}\rightarrow c\epsilon_{-}.
\end{aligned}
\end{equation}
All quantities not displayed in these expressions are held fixed. At
the level of the Lagrangian density, the two prescriptions give
\begin{equation}
\mathcal{L}_{\mathrm{rel}}
\xrightarrow{\mathrm{I}}
\mathcal{L}_{\mathrm{hyb}}
+c^2\mathcal{L}_{\mathrm{mag}},
\qquad
\mathcal{L}_{\mathrm{rel}}
\xrightarrow{\mathrm{II}}
c^{-2}\mathcal{L}_{\mathrm{hyb}}
+\mathcal{L}_{\mathrm{mag}}.
\end{equation}

This decomposition is not specific to the present vector multiplet. As
discussed in \cite{Ergec:2026baz}, the same reasoning applies generally
to any theory whose contraction decomposes the parent Lagrangian into
distinct, individually consistent sectors. For supersymmetric theories,
these sectors are realized by either the contracted multiplet or one of
its truncated submultiplets.

Multiplying the result of the second prescription by the overall factor
$c^2$, both scalings give
$\mathcal{L}_{\mathrm{hyb}}+c^2\mathcal{L}_{\mathrm{mag}}$ and therefore
yield the same Carroll limit as $c\rightarrow0$. The resulting off-shell
multiplet is described by $\mathcal{L}_{\mathrm{hyb}}$ and the
transformations
\begin{equation}
\label{eq:2p2-vector-multiplet}
\begin{aligned}
\delta A_{0}
&=-\frac{1}{2}\bar\epsilon_{+}\gamma_{0}\lambda_{-},
\\[2pt]
\delta\sigma
&=\frac{1}{2}\left(
\bar\epsilon_{+}\gamma_{0}\lambda_{+}
-\bar\epsilon_{-}\gamma_{0}\lambda_{-}
\right),
\\[2pt]
\delta A_{a}
&=-\frac{1}{2}\left(
\bar\epsilon_{+}\gamma_{a}\lambda_{+}
+\bar\epsilon_{-}\gamma_{a}\lambda_{-}
\right),
\\[2pt]
\delta\lambda_{+}
&=
\left(
-\frac{1}{4}\gamma^{ab}F_{ab}
-\frac{1}{2}\gamma_{0}\gamma^{a}F_{ya}
\right)\epsilon_{+}
\\
&\quad+
\left(
-\frac{1}{2}\gamma^{0a}F_{0a}
-\frac{1}{2}F_{y0}
+\frac{1}{2}D\gamma_{0}
\right)\epsilon_{-},
\\[2pt]
\delta\lambda_{-}
&=
\left(
-\frac{1}{2}\gamma^{0a}F_{0a}
+\frac{1}{2}F_{y0}
-\frac{1}{2}D\gamma_{0}
\right)\epsilon_{+},
\\[2pt]
\delta D
&=
\frac{1}{2}\bar\epsilon_{+}
\left(
\partial_{0}\lambda_{+}
+\gamma_{0}\gamma^{a}\partial_{a}\lambda_{-}
+\partial_{y}\lambda_{-}
\right)
-\frac{1}{2}\bar\epsilon_{-}\partial_{0}\lambda_{-}.
\end{aligned}
\end{equation}
The sector $\mathcal{L}_{\text{hyb}}$ has hybrid dynamics: its bosonic
sector is of electric type, whereas its fermionic sector retains
magnetic spatial dynamics. The transformations close off shell as
\begin{equation}
[\delta^{(1)},\delta^{(2)}]\Phi
=
\xi^0\partial_0\Phi
+\xi^a\partial_a\Phi
+\xi^y\partial_y\Phi
+\delta_{\Lambda}^{\mathrm{gauge}}\Phi,
\end{equation}
where the translation parameters are
\begin{equation}
\begin{aligned}
\xi^0
&=\bar{\epsilon}^{[(1)}_{+}\gamma^0\epsilon^{(2)]}_{-},
\\
\xi^a
&=\frac{1}{2}\bar{\epsilon}^{[(1)}_{+}
\gamma^a\epsilon^{(2)]}_{+},
\\
\xi^y
&=-\frac{1}{2}\bar{\epsilon}^{[(1)}_{+}
\gamma^0\epsilon^{(2)]}_{+}.
\end{aligned}
\end{equation}
The gauge parameter is
\begin{equation}
\Lambda=-\xi^0A_0-\xi^aA_a-\xi^y\sigma,
\end{equation}
with
\begin{equation}
\delta_{\Lambda}^{\mathrm{gauge}}A_M=\partial_M\Lambda,
\qquad
\delta_{\Lambda}^{\mathrm{gauge}}\lambda_{\pm}
=
\delta_{\Lambda}^{\mathrm{gauge}}D=0.
\end{equation}

A magnetic-type sector can nevertheless be isolated as a consistent
half-supersymmetric truncation of the hybrid multiplet. This is achieved
by restricting the supersymmetry parameter and the fields according to
\begin{equation}
\epsilon_-=0,
\qquad
A_0=0,
\qquad
\lambda_-=0,
\qquad
D=0,
\end{equation}
together with the time-independence constraints
\begin{equation}
\partial_0\sigma=0,
\qquad
\partial_0A_a=0,
\qquad
\partial_0\lambda_+=0,
\end{equation}
which characterize the magnetic Carroll sector. Under these conditions,
$F_{0a}=F_{0y}=0$ and $\mathcal{L}_{\text{hyb}}$ vanishes, so that the
surviving Lagrangian is precisely $\mathcal{L}_{\text{mag}}$. The
surviving supersymmetry transformations are
\begin{equation}
\begin{aligned}
\delta\sigma
&=\frac{1}{2}\bar{\epsilon}_{+}\gamma_0\lambda_{+},
\\
\delta A_a
&=-\frac{1}{2}\bar{\epsilon}_{+}\gamma_a\lambda_{+},
\\
\delta\lambda_{+}
&=
\left(
-\frac{1}{4}\gamma^{ab}F_{ab}
-\frac{1}{2}\gamma_0\gamma^aF_{ya}
\right)\epsilon_{+}.
\end{aligned}
\end{equation}
These transformations preserve the time-independence constraints.
Indeed, $\delta(\partial_0\sigma)$ and $\delta(\partial_0A_a)$ are
proportional to $\partial_0\lambda_+$, while
$\delta(\partial_0\lambda_+)$ depends on $\partial_0F_{ab}$ and
$\partial_0F_{ya}$, both of which vanish as a consequence of the
constraints. Thus,
\begin{equation}
\delta(\partial_0\sigma)
=
\delta(\partial_0A_a)
=
\delta(\partial_0\lambda_+)
=
0.
\end{equation}
On this constrained field space, $\mathcal{L}_{\text{mag}}$ is invariant
under the truncated supersymmetry and Carroll boosts, thereby defining
a consistent constrained magnetic Carroll multiplet. Thus, the direct
contraction yields the hybrid off-shell theory, while
$\mathcal{L}_{\text{mag}}$ arises as a consistent supersymmetric
magnetic truncation.

%%%%%%%%%%%%%%%%%%%%%%%%%%%%%%%%%%%%%%%%%%%%%%%%%%%%%%%%%%%%%%%%%%%%%%%%%%%%%%%%
\textit{Conclusion}--
%%%%%%%%%%%%%%%%%%%%%%%%%%%%%%%%%%%%%%%%%%%%%%%%%%%%%%%%%%%%%%%%%%%%%%%%%%%%%%%%
In this work, we realize the $2+1$ dimensional twisted $\mathcal{N}=2$ Carroll superalgebra from a $2+2$ dimensional parent structure. The higher dimensional parent theory also allows us to interpret consistent truncations in terms of chiral super- $B M S_4$-type algebras, which may be glued together through the $R$-symmetry. In addition, we construct consistent off-shell hybrid and magnetic theories realizing this Carrollian symmetry and clarify the additional consistency conditions required in the on-shell case. We further show that the magnetic sector does not, in general, arise from a naive contraction, and that the hybrid contraction can instead serve as a useful guide for identifying the corresponding magnetic theory.\\

Moreover, string theory signals structures that still lie beyond our current understanding. One of the clearest examples is the AdS/CFT correspondence. In this context, it is natural to expect that flat-space holography may admit a Carrollian interpretation through its connection to $BMS$ structures, although much remains to be clarified. In particular, one must first understand how the contraction is realized as a generalized symmetry on the field-theory side, and develop a systematic way to extract the corresponding electric and magnetic theories. On the gravity side, the same contraction acquires a geometric interpretation, in which the spacetime manifold becomes degenerate while remaining locally consistent. These two descriptions are expected to meet within the holographic picture.

As mentioned, the Carrollian idea has appeared in several distinct forms. Geometrically, the Carrollian limit makes the spacetime manifold degenerate. At the level of field theory, it also may be realized through a degenerate Clifford structure, often referred to as the Carrollian Clifford algebra \cite{Bergshoeff:2023vfd, Grumiller:2025rtm}, which underlies the construction of Carrollian spinors and Carroll--Lorentz scalar field theories. In the present work, however, we interpret the Carrollian structure as a restriction to a consistent sector of the parent theory. From this viewpoint, it may be regarded as a generalized symmetry inherited from the parent theory, and the Carrollian limit should therefore be understood as a consistent contraction that may reveal hidden aspects of CFTs. Since this contraction is implemented at the level of a consistent algebra, and the multiplet structures realize this symmetry as the building blocks of the theory, the apparent divergences that arise in some cases may point not to an inconsistency but rather to a missing step or hidden ingredient in the contraction procedure. We also expect the geometric side to be significantly more difficult, since the global structure becomes degenerate, whereas the standard formulation of general relativity is built on a nondegenerate geometry in which space and time are treated on equal footing. This suggests that a different framework may be required for the geometry, even though non-relativistic versions of gravity already admit physically meaningful realizations. 

As future work, given the degeneracy of Lorentzian geometry in the Carroll limit, it would be important to construct a twisted Carroll supergravity theory, building on the recent Carroll supergravity construction of \cite{Henneaux:2026dfc}, and to investigate whether the Carrollian theories constructed here remain anomaly-free at the quantum level. Moreover, the analysis of \cite{Ergec:2026baz} can be extended to other contracted theories. It suggests that an İnönü–Wigner contraction decomposes a parent theory into distinct, individually consistent sectors, a possibility that deserves systematic investigation.

%%%%%%%%%%%%%%%%%%%%%%%%%%%%%%%%%%%%%%%%%%%%%%%%%%%%%%%%%%%%%%%%%%%%%%%%%%%%%%%%
\textit{Acknowledgements}--
%%%%%%%%%%%%%%%%%%%%%%%%%%%%%%%%%%%%%%%%%%%%%%%%%%%%%%%%%%%%%%%%%%%%%%%%%%%%%%%%
We thank the theory group at Istanbul Technical University, especially Mehmet Ozkan, Oguzhan Kasikci, Ilayda Bulunur, Mustafa Salih Zog, and Enes Bal, for valuable discussions.
%%%%%%%%%%%%%%%%%%%%%%%%%%%%%%%%%%%%%%%%%%%%%

\bibliography{ref}

\section*{SUPPLEMENTAL MATERIAL \\ (APPENDICES)}

\section*{APPENDIX A: Twisting Procedure for Supersymmetric Theories}
Twisting either the algebraic or field-theoretic description is straightforward. These structures can be obtained from ordinary theories by analytic continuation. We first express the entire structure in terms of Majorana spinors and then analytically continue one of the Majorana supercharges, say,
\begin{equation}
    Q_2 \longrightarrow iQ_2 , \quad \bar Q_2 \longrightarrow i\bar Q_2 
\end{equation}
At the level of the supersymmetry transformations, this is
accompanied by
\begin{equation}
    \epsilon_2 \longrightarrow i\epsilon_2, \quad  \bar \epsilon_2 \longrightarrow i \bar \epsilon_2
\end{equation}
for the corresponding supersymmetry parameter. One must then identify suitable complex rescalings of the remaining
generators and component fields that eliminate all explicit factors of
$i$ from the algebra, the field-theory description, and the multiplet
transformation rules, without applying any further complex rescalings
to the Majorana supercharges $Q_i, \bar Q_i$ or their corresponding parameters
$\epsilon_i, \bar \epsilon_i $.

These invertible redefinitions preserve the graded Jacobi identities and the closure of the multiplet. The continuation preserves $\bar{Q}_2=Q_2^{T}C$ but selects a different real form, in which the transformed supercharges and supersymmetry parameters are again taken to be Majorana. Accordingly, the internal metric in the closure relation changes from $\delta^{ij}$ to $\eta^{ij}$.

\begin{equation}
\begin{gathered}
\left[\delta^{(1)},\delta^{(2)}\right]\Phi
=
\frac{1}{2}\delta^{ij}
\bar{\epsilon}_{i}^{(2)}\gamma^\mu
\epsilon_{j}^{(1)}\partial_\mu\Phi
\\[2pt]
\delta^{ij}\longrightarrow\eta^{ij}
\\[2pt]
\left[\delta^{(1)},\delta^{(2)}\right]\Phi
=
\frac{1}{2}\eta^{ij}
\bar{\epsilon}_{i}^{(2)}\gamma^\mu
\epsilon_{j}^{(1)}\partial_\mu\Phi .
\end{gathered}
\end{equation}
For two supercharges, $\eta^{ij}=\operatorname{diag}(1,-1)$ up to conventions. For example, consider the following ordinary $\mathcal{N}=2$ vector multiplet in $2+1$ dimensions
\begin{equation}
    \begin{aligned}
\delta \rho & =\varepsilon^{i j} \bar{\epsilon}_i \lambda_j, \\
\delta \lambda_i & =-\frac{1}{2} \gamma^\mu \epsilon_i V_\mu-\frac{1}{2} \varepsilon^{i j} \epsilon_j D-\frac{1}{4} \gamma^\mu \varepsilon^{i j} \epsilon_j \partial_\mu \rho, \\
\delta D & =\frac{1}{2} \varepsilon^{i j} \bar{\epsilon}_i \gamma^\mu \partial_\mu \lambda_j, \\
\delta V_\mu & =\frac{1}{2} \delta^{i j} \bar{\epsilon}_i \gamma_\mu^v \partial_\nu \lambda_j .
\end{aligned}
\end{equation}

with $\varepsilon^{12}=1$. We first perform the analytic continuation

\begin{equation}
\begin{aligned}
    \epsilon_2\longrightarrow i\epsilon_2,
    \qquad
    \bar{\epsilon}_2\longrightarrow i\bar{\epsilon}_2 .
    \end{aligned}
\end{equation}

The supersymmetry transformations then become

\begin{equation}
\begin{aligned}
\delta\rho
&=
\bar{\epsilon}_1\lambda_2
-i\bar{\epsilon}_2\lambda_1,
\\
\delta\lambda_1
&=
-\frac{1}{2}\gamma^\mu\epsilon_1V_\mu
-\frac{i}{2}\epsilon_2D
-\frac{i}{4}\gamma^\mu\epsilon_2\partial_\mu\rho,
\\
\delta\lambda_2
&=
-\frac{i}{2}\gamma^\mu\epsilon_2V_\mu
+\frac{1}{2}\epsilon_1D
+\frac{1}{4}\gamma^\mu\epsilon_1\partial_\mu\rho,
\\
\delta D
&=
\frac{1}{2}\bar{\epsilon}_1\gamma^\mu
\partial_\mu\lambda_2
-\frac{i}{2}\bar{\epsilon}_2\gamma^\mu
\partial_\mu\lambda_1,
\\
\delta V_\mu
&=
\frac{1}{2}\bar{\epsilon}_1
\gamma_\mu{}^\nu\partial_\nu\lambda_1
+\frac{i}{2}\bar{\epsilon}_2
\gamma_\mu{}^\nu\partial_\nu\lambda_2 .
\end{aligned}
\end{equation}

To eliminate the remaining explicit factors of $i$, while leaving the
supersymmetry parameters unchanged, we perform the complex field
redefinitions

\begin{equation}
\begin{aligned}
        \lambda_1\rightarrow-i\lambda_1, \quad  \bar\lambda_1\rightarrow-i \bar\lambda_1,
    \quad
    V_\mu\rightarrow iV_\mu,
    \end{aligned}
\end{equation}

with $\rho$, $D$, and $\lambda_2$ left unchanged, the resulting transformations take the twisted form

\begin{equation}
\begin{aligned}
\delta\rho
&=
\bar{\epsilon}_1\lambda_2
-\bar{\epsilon}_2\lambda_1,
\\
\delta\lambda_1
&=
\frac{1}{2}\gamma^\mu\epsilon_1V_\mu
+\frac{1}{2}\epsilon_2D
+\frac{1}{4}\gamma^\mu\epsilon_2\partial_\mu\rho,
\\
\delta\lambda_2
&=
\frac{1}{2}\gamma^\mu\epsilon_2V_\mu
+\frac{1}{2}\epsilon_1D
+\frac{1}{4}\gamma^\mu\epsilon_1\partial_\mu\rho,
\\
\delta D
&=
\frac{1}{2}\bar{\epsilon}_1\gamma^\mu
\partial_\mu\lambda_2
-\frac{1}{2}\bar{\epsilon}_2\gamma^\mu
\partial_\mu\lambda_1,
\\
\delta V_\mu
&=
-\frac{1}{2}\bar{\epsilon}_1
\gamma_\mu{}^\nu\partial_\nu\lambda_1
+\frac{1}{2}\bar{\epsilon}_2
\gamma_\mu{}^\nu\partial_\nu\lambda_2 .
\end{aligned}
\end{equation}

For the algebraic construction, the same procedure applies in the superconformal case. It is important to note that the $R$-symmetry generator must also be rescaled by $i$ to remove the remaining factors of $i$ from the algebra. This changes the elliptic $SO(2)$ rotations into hyperbolic $SO(1,1)$ rotations.

\section*{APPENDIX B: Off-Shell Mass Deformation of the Twisted Carrollian
$\mathcal{N}=2$ Scalar Multiplet}

The first scalar realization introduced in the main text admits an off-shell mass deformation. Its complete transformation rules are
\begin{equation}
\label{app:massive-scalar}
\begin{gathered}
\delta A_1=\bar{\epsilon}_{+}\chi_{-},
\qquad
\delta A_2=\bar{\epsilon}_{+}\gamma_0\chi_{+},
\\
\delta\chi_{+}
=
\left(
-\gamma_0\gamma^a\partial_a A_2
-mA_2
+\frac{1}{2}\gamma_0F_2
\right)\epsilon_{+}
-2(\partial_0A_2)\epsilon_{-},
\\
\delta\chi_{-}
=
\left(
\gamma^a\partial_a A_1
-\frac{1}{2}F_1
-m\gamma_0A_1
\right)\epsilon_{+}
+2\gamma^0(\partial_0A_1)\epsilon_{-},
\\
\delta F_1
=
-2\bar{\epsilon}_{+}
\left(
\gamma^a\partial_a\chi_{-}
-m\gamma_0\chi_{-}
\right)
-4\bar{\epsilon}_{-}\gamma^0\partial_0\chi_{-},
\\
\delta F_2
=
2\bar{\epsilon}_{+}
\left(
\gamma_0\gamma^a\partial_a\chi_{+}
-m\chi_{+}
\right)
-4\bar{\epsilon}_{-}\partial_0\chi_{+}.
\end{gathered}
\end{equation}
These transformations close off shell according to
\begin{equation}
\begin{aligned}
[\delta^{(1)},\delta^{(2)}]\Phi
={}&
-2\left(
\bar{\epsilon}_{+}^{(1)}\gamma^0\epsilon_{-}^{(2)}
+
\bar{\epsilon}_{-}^{(1)}\gamma^0\epsilon_{+}^{(2)}
\right)\partial_0\Phi
\\
&-2\bar{\epsilon}_{+}^{(1)}
\gamma^a\epsilon_{+}^{(2)}\partial_a\Phi
+2\bar{\epsilon}_{+}^{(1)}
\gamma_0\epsilon_{+}^{(2)}Z\Phi,
\end{aligned}
\end{equation}
where \(Z\) is the central-charge generator, acting on the multiplet as
\begin{equation}
\begin{aligned}
    Z\Phi=m\Phi.
\end{aligned}
\end{equation}
%%%%%%%%%%%%%%%%%%%%%%%%%%%%%%%%%%%%%%%%%%%%%%%%%%%%%%%%%%%%%%%%%%%%%%%%%%%%%%%%
\section*{ APPENDIX C: $2+2$ dimensional superconformal Carrolian algebra }
%%%%%%%%%%%%%%%%%%%%%%%%%%%%%%%%%%%%%%%%%%%%%%%%%%%%%%%%%%%%%%%%%%%%%%%%%%%%%%%%

Applying the Carroll scaling in Eq. \eqref{Carrollian scaling of 2+2 d superconformalalgebra}
and taking $c\rightarrow 0$ gives the following nonvanishing brackets.

Bosonic sector $X_a\in\{C_a,P_a,B_a,K_a\}$
\begin{equation}
\begin{aligned}
[M_{ab},M_{cd}]
&=4\delta_{[a[c}M_{d]b]}, \quad &
[C_a,B_b]=\delta_{ab}B,
\\
[M_{ab},X_c]
&=2\delta_{c[b}X_{a]}, \quad &[H,K_a]=-2C_a,
\\[2pt]
[C_a,P_b]&=\delta_{ab}H,
&
[C_a,K_b]=\delta_{ab}K,
\\
[B_a,B_b]&=M_{ab},
&
[B,B_a]=C_a,
\\
[B,Z]&=-H,
&
[B,\widetilde Z]=-K,
\\[2pt]
[B_a,P_b]&=-\delta_{ab}Z,
&
[B_a,Z]=-P_a,
\\
[B_a,K_b]&=-\delta_{ab}\widetilde Z,
&
[B_a,\widetilde Z]=-K_a,
\\[2pt]
[P_a,K_b]
&=2\bigl(\delta_{ab}D-M_{ab}\bigr),
&
[P_a,K]=2C_a,
\\
[P_a,\widetilde Z]&=-2B_a,
&
[H,\widetilde Z]=-2B,
\\
[Z,K_a]&=2B_a,
&
[Z,K]=2B,
\\[2pt]
[P_a,D]&=P_a,
&
[H,D]=H,
\\
[K_a,D]&=-K_a,
&
[K,D]=-K,
\\
[Z,D]&=Z,
&
[\widetilde Z,D]=-\widetilde Z.\\
 [Z,\widetilde Z]&=-2D.
\end{aligned}
\label{app:full-contracted-bosonic}
\end{equation}

Mixed sector $\Psi^\pm\in\{Q^\pm,S^\pm\}$ 

\begin{equation}
\begin{aligned}
[M_{ab},\Psi^\pm]
&=-\frac12\gamma_{ab}\Psi^\pm,
&
[B_a,\Psi^\pm]
&=\pm\frac12\gamma_{a0}\Psi^\pm,
\\[2pt]
[P_a,S^\pm]&=\gamma_aQ^\pm,
&
[K_a,Q^\pm]&=-\gamma_aS^\pm,
\\
[H,S^+]&=\gamma_0Q^-,
&
[K,Q^+]&=-\gamma_0S^-,
\\
[Z,S^\pm]&=\mp\gamma_0Q^\pm,
&
[\widetilde Z,Q^\pm]&=\pm\gamma_0S^\pm,
\\[2pt]
[C_a,Q^+]&=-\frac12\gamma_{0a}Q^-,
&
[C_a,S^+]&=-\frac12\gamma_{0a}S^-,
\\
[B,Q^+]&=\frac12Q^-,
&
[B,S^+]&=\frac12S^-,
\\[2pt]
[D,Q^\pm]&=-\frac12Q^\pm,
&
[D,S^\pm]&=\frac12S^\pm,
\\
[A,Q^+]&=3\gamma_0Q^-,
&
[A,S^+]&=-3\gamma_0S^-.
\end{aligned}
\label{app:full-contracted-mixed}
\end{equation}

Fermionic sector

\begin{equation}
\begin{aligned}
\{Q^+,Q^+\}
&=
2\gamma^aC_3\,P_a
-2\gamma^0C_3\,Z,
\\
\{Q^+,Q^-\}
&=
2\gamma^0C_3\,H,
\\[2pt]
\{S^+,S^+\}
&=
-2\gamma^aC_3\,K_a
+2\gamma^0C_3\,\widetilde Z,
\\
\{S^+,S^-\}
&=
-2\gamma^0C_3\,K,
\\[2pt]
\{S^+,Q^+\}
&=
2C_3D
+\gamma^{ab}C_3\,M_{ab}
+2\gamma^{0a}C_3\,B_a,
\\
\{S^+,Q^-\}
&=
-2C_3B
-\gamma^0C_3\,A
+2\gamma^{0a}C_3\,C_a,
\\
\{S^-,Q^+\}
&=
2C_3B
+\gamma^0C_3\,A
+2\gamma^{0a}C_3\,C_a.
\end{aligned}
\label{app:full-contracted-fermionic}
\end{equation}

All (anti)commutators not displayed above vanish. This is the full contracted
algebra written in a $2+1$ dimensional basis. The Type A and Type B algebras
discussed in the main text follow from subsequent consistent truncations of
this structure.

\end{document}